\documentclass[twocolumn,pre,aps,nofootinbib,superscriptaddress,showpacs,preprintnumbers]{revtex4-1}

\usepackage{graphicx,amsmath}
\graphicspath{{./fig/}}

\usepackage{bm}

\begin{document}

\preprint{NORDITA-2011-6}

%
\newcommand{\EQ}{\begin{equation}}
\newcommand{\EN}{\end{equation}}
\newcommand{\EQA}{\begin{eqnarray}}
\newcommand{\ENA}{\end{eqnarray}}
\newcommand{\eq}[1]{(\ref{#1})}
\newcommand{\Eq}[1]{Eq.~(\ref{#1})}
\newcommand{\Eqs}[2]{Eqs.~(\ref{#1}) and~(\ref{#2})}
\newcommand{\eqs}[2]{(\ref{#1}) and~(\ref{#2})}
\newcommand{\Eqss}[2]{Eqs.~(\ref{#1})--(\ref{#2})}
\newcommand{\eqss}[2]{(\ref{#1})--(\ref{#2})}
\newcommand{\Section}[1]{Sec.\,\ref{#1}}
\newcommand{\Sec}[1]{\S\,\ref{#1}}
\newcommand{\App}[1]{Appendix~\ref{#1}}
\newcommand{\Fig}[1]{Fig.~\ref{#1}}
\newcommand{\Tab}[1]{Table~\ref{#1}}
\newcommand{\Figs}[2]{Figures~\ref{#1} and \ref{#2}}
\newcommand{\Tabs}[2]{Tables~\ref{#1} and \ref{#2}}
\newcommand{\bra}[1]{\langle #1\rangle}
\newcommand{\bbra}[1]{\left\langle #1\right\rangle}
\newcommand{\mean}[1]{\overline #1}
\newcommand{\meanEMF}{\overline{\mbox{\boldmath ${\cal E}$}} {}}
\newcommand{\meanFF}{\overline{\mbox{\boldmath ${\cal F}$}} {}}
\newcommand{\meanB}{\overline{B}}
\newcommand{\meanF}{\overline{\cal F}}
\newcommand{\meanJ}{\overline{J}}
\newcommand{\meanU}{\overline{U}}
\newcommand{\meanT}{\overline{T}}
\newcommand{\meanrho}{\overline{\rho}}
\newcommand{\UU}{{\bm U}}
\newcommand{\UUp}{{\bm U}_{\rm p}}
\newcommand{\Lg}{L_{\rm g}}
\newcommand{\rhop}{\rho_{\rm p}}
\newcommand{\SSS}{{\sf S}}
\newcommand{\SSSS}{\mbox{\boldmath ${\sf S}$} {}}
\newcommand{\meanAA}{\overline{\mbox{\boldmath $A$}}}
\newcommand{\meanBB}{\overline{\mbox{\boldmath $B$}}}
\newcommand{\meanUU}{\overline{\mbox{\boldmath $U$}}}
\newcommand{\meanJJ}{\overline{\mbox{\boldmath $J$}}}
\newcommand{\meanEE}{\overline{\mbox{\boldmath $E$}}}
\newcommand{\meanuu}{\overline{\mbox{\boldmath $u$}}}
\newcommand{\meanAB}{\overline{\mbox{\boldmath $A\cdot B$}}}
\newcommand{\meanAoBo}{\overline{\mbox{\boldmath $A_0\cdot B_0$}}}
\newcommand{\ff}{\bm{f}}
\newcommand{\xx}{\bm{x}}
\newcommand{\kk}{\bm{k}}
\newcommand{\ii}{\mathrm{i}}
\newcommand{\eee}{\bm{e}}
\newcommand{\meanApoBpo}{\overline{\mbox{\boldmath $A'_0\cdot B'_0$}}}
\newcommand{\meanApBp}{\overline{\mbox{\boldmath $A'\cdot B'$}}}
\newcommand{\meanuxB}{\overline{\mbox{\boldmath $\delta u\times \delta B$}}}
\newcommand{\chk}[1]{[{\em check: #1}]}
\newcommand{\p}{\partial}
\newcommand{\xder}[1]{\frac{\partial #1}{\partial x}}
\newcommand{\yder}[1]{\frac{\partial #1}{\partial y}}
\newcommand{\zder}[1]{\frac{\partial #1}{\partial z}}
\newcommand{\xdertwo}[1]{\frac{\partial^2 #1}{\partial x^2}}
\newcommand{\xderj}[2]{\frac{\partial #1}{\partial x_{#2}}}
\newcommand{\timeder}[1]{\frac{\partial #1}{\partial t}}
\newcommand{\bec}[1]{\mbox{\boldmath $ #1$}}
\newcommand{\nab}{\mbox{\boldmath $\nabla$} {}}
%
%
\def\const{\rm const}
\def\dd{{\rm d}}
\def\DD{{\rm D}}
\def\tauf{\tau_{\rm f}}
\def\tauk{\tau_{\rm k}}
\def\taup{\tau_{\rm p}}
\def\Dc{D_{\rm c}}
\def\Dt{D_{\rm t}}
\def\DT{D_{\rm T}}
\def\tauc{\tau_{\rm c}}
\def\taueff{\tau_{\rm eff}}
\def\uoneD{{v'}}
\def\urms{u_{\rm rms}}
\def\cs{c_{\rm s}}
\def\kf{k_{\rm f}}
\def\sT{s_{\rm T}}
\def\sL{s_{\rm L}}
\def\vF{s_{\rm T}}

\def\Rey{\mbox{\rm Re}}
\def\Nu{\mbox{\rm Nu}}
\def\Bi{\mbox{\rm Bi}}
\def\Pe{\mbox{\rm Pe}}
\def\Sc{\mbox{\rm Sc}}
\def\St{\mbox{\rm St}}

%
%
\newcommand{\yan}[3]{, Astron. Nachr. {\bf #2}, #3 (#1).}
\newcommand{\yact}[3]{, Acta Astron. {\bf #2}, #3 (#1).}
\newcommand{\yana}[3]{, Astron. Astrophys. {\bf #2}, #3 (#1).}
\newcommand{\yanas}[3]{, Astron. Astrophys. Suppl. {\bf #2}, #3 (#1).}
\newcommand{\yanal}[3]{, Astron. Astrophys. Lett. {\bf #2}, #3 (#1).}
\newcommand{\yass}[3]{, Astrophys. Spa. Sci. {\bf #2}, #3 (#1).}
\newcommand{\ysci}[3]{, Science {\bf #2}, #3 (#1).}
\newcommand{\ysph}[3]{, Solar Phys. {\bf #2}, #3 (#1).}
\newcommand{\yjetp}[3]{, Sov. Phys. JETP {\bf #2}, #3 (#1).}
\newcommand{\yspd}[3]{, Sov. Phys. Dokl. {\bf #2}, #3 (#1).}
\newcommand{\ysov}[3]{, Sov. Astron. {\bf #2}, #3 (#1).}
\newcommand{\ysovl}[3]{, Sov. Astron. Letters {\bf #2}, #3 (#1).}
\newcommand{\ymn}[3]{, Monthly Notices Roy. Astron. Soc. {\bf #2}, #3 (#1).}
\newcommand{\yqjras}[3]{, Quart. J. Roy. Astron. Soc. {\bf #2}, #3 (#1).}
\newcommand{\ynat}[3]{, Nature {\bf #2}, #3 (#1).}
\newcommand{\sjfm}[2]{, J. Fluid Mech., submitted (#1).}
\newcommand{\pjfm}[2]{, J. Fluid Mech., in press (#1).}
\newcommand{\yjfm}[3]{, J. Fluid Mech. {\bf #2}, #3 (#1).}
\newcommand{\ypepi}[3]{, Phys. Earth Planet. Int. {\bf #2}, #3 (#1).}
\newcommand{\ypr}[3]{, Phys.\ Rev.\ {\bf #2}, #3 (#1).}
\newcommand{\yprl}[3]{, Phys.\ Rev.\ Lett.\ {\bf #2}, #3 (#1).}
\newcommand{\yepl}[3]{, Europhys. Lett. {\bf #2}, #3 (#1).}
\newcommand{\pcsf}[2]{, Chaos, Solitons \& Fractals, in press (#1).}
\newcommand{\ycsf}[3]{, Chaos, Solitons \& Fractals{\bf #2}, #3 (#1).}
\newcommand{\yprs}[3]{, Proc. Roy. Soc. Lond. {\bf #2}, #3 (#1).}
\newcommand{\yptrs}[3]{, Phil. Trans. Roy. Soc. {\bf #2}, #3 (#1).}
\newcommand{\yjcp}[3]{, J. Comp. Phys. {\bf #2}, #3 (#1).}
\newcommand{\yjgr}[3]{, J. Geophys. Res. {\bf #2}, #3 (#1).}
\newcommand{\ygrl}[3]{, Geophys. Res. Lett. {\bf #2}, #3 (#1).}
\newcommand{\yobs}[3]{, Observatory {\bf #2}, #3 (#1).}
\newcommand{\yaj}[3]{, Astronom. J. {\bf #2}, #3 (#1).}
\newcommand{\yapj}[3]{, Astrophys. J. {\bf #2}, #3 (#1).}
\newcommand{\yapjs}[3]{, Astrophys. J. Suppl. {\bf #2}, #3 (#1).}
\newcommand{\yapjl}[3]{, Astrophys. J. {\bf #2}, #3 (#1).}
\newcommand{\ypp}[3]{, Phys. Plasmas {\bf #2}, #3 (#1).}
\newcommand{\ypasj}[3]{, Publ. Astron. Soc. Japan {\bf #2}, #3 (#1).}
\newcommand{\ypac}[3]{, Publ. Astron. Soc. Pacific {\bf #2}, #3 (#1).}
\newcommand{\yannr}[3]{, Ann. Rev. Astron. Astrophys. {\bf #2}, #3 (#1).}
\newcommand{\yanar}[3]{, Astron. Astrophys. Rev. {\bf #2}, #3 (#1).}
\newcommand{\yanf}[3]{, Ann. Rev. Fluid Dyn. {\bf #2}, #3 (#1).}
\newcommand{\ypf}[3]{, Phys. Fluids {\bf #2}, #3 (#1).}
\newcommand{\yphy}[3]{, Physica {\bf #2}, #3 (#1).}
\newcommand{\ygafd}[3]{, Geophys. Astrophys. Fluid Dynam. {\bf #2}, #3 (#1).}
\newcommand{\yzfa}[3]{, Zeitschr. f. Astrophys. {\bf #2}, #3 (#1).}
\newcommand{\yptp}[3]{, Progr. Theor. Phys. {\bf #2}, #3 (#1).}
\newcommand{\yjour}[4]{, #2 {\bf #3}, #4 (#1).}
\newcommand{\pjour}[3]{, #2, in press (#1).}
\newcommand{\sjour}[3]{, #2, submitted (#1).}
\newcommand{\yprep}[2]{, #2, preprint (#1).}
\newcommand{\pproc}[3]{, (ed. #2), #3 (#1) (to appear).}
\newcommand{\yproc}[4]{, (ed. #3), pp. #2. #4 (#1).}
\newcommand{\ybook}[3]{, {\em #2}. #3 (#1).}

\title{Detection of turbulent thermal diffusion of particles in numerical simulations}
\author{Nils Erland L. Haugen}
\email{Nils.E.Haugen@sintef.no}
\affiliation{SINTEF Energy Research, N-7034 Trondheim, Norway}

\author{Nathan Kleeorin}
\email{nat@bgu.ac.il}
\author{Igor Rogachevskii}
\email{gary@bgu.ac.il}
\affiliation{Department of Mechanical Engineering,
Ben-Gurion University of the Negev, P.O.Box 653, Beer-Sheva 84105,  Israel}

\author{Axel Brandenburg}
\email{brandenb@nordita.org}
\affiliation{NORDITA, AlbaNova University Center,
Roslagstullsbacken 23, SE 10691 Stockholm, Sweden}
\affiliation{Department of Astronomy,
Stockholm University, SE 10691 Stockholm, Sweden}

\date{\today,~ $ $Revision: 1.177 $ $}

\begin{abstract}
The phenomenon of turbulent thermal diffusion in
temperature-stratified turbulence causing
a non-diffusive turbulent flux of inertial and non-inertial
particles in the direction of the turbulent heat flux is
found using direct numerical simulations (DNS).
In simulations with and without gravity,
this phenomenon is found to cause a peak in the particle
number density around the minimum of the mean fluid
temperature for Stokes numbers less than 1,
where the Stokes number is the ratio of particle Stokes time
to turbulent Kolmogorov time at the viscous scale.
Turbulent thermal diffusion causes the formation of large-scale inhomogeneities
in the spatial distribution of inertial particles.
The strength of this effect is maximum for Stokes numbers around unity,
and decreases again for larger values.
The dynamics of inertial particles is studied using
Lagrangian modelling in forced temperature-stratified
turbulence, whereas non-inertial particles and the fluid
are described using DNS in an Eulerian framework.

\end{abstract}
\pacs{
47.27.tb, 
47.27.T-, 
47.55.Hd, 
47.55.Kf  
}
\maketitle

\section{Introduction}

Transport and mixing of small particles (aerosols and droplets)
in turbulent fluid flow is of
fundamental importance in a large variety of applications
(environmental sciences, physics of the atmosphere
and meteorology, industrial turbulent flows and
turbulent combustion; see, e.g.,
\cite{CSA80,BLA97,CST11,WA00,BE10,S03,BH03,KPE07}).
There are also astrophysical applications,
in particular in the context of
protoplanetary accretion discs \cite{SA72,TA09,HB98,JAB04}.

Numerous laboratory \cite{WA00,AC02,WH05,SA08,XB08,SSA08}
and numerical \cite{BL04,CC05,CG06,YG07,AC08,BB07,BB10}
experiments as well as observations
in atmospheric \cite{S03,KPE07,VY00,SGG10} and
astrophysical \cite{SA72,TA09,HB98,JAB04,JY07,Johansen07} turbulent
flows have shown different kinds of large-scale and small-scale
long-living inhomogeneities (clusters) in the spatial
distribution of particles.
It is well known that turbulent diffusion causes destruction of large-scale inhomogeneities in the spatial
distributions of particles.
But how can we explain the opposite process resulting in a formation
of large-scale clusters of particles?

One of the mechanisms of
formation of particle inhomogeneities
in temperature-stratified turbulence
is the phenomenon of turbulent {\it thermal} diffusion \cite{elperin_etal96}.
This effect consists of a turbulent
non-diffusive flux of inertial particles in the
direction of the turbulent heat flux, so that
particles are accumulated in the vicinity of the
mean temperature minimum.
The particular form of the flow field does not play any role
in this effect. It is a purely collective phenomenon
caused by temperature stratified turbulence resulting in
a pumping effect, i.e.,
appearance of the non-zero mean effective velocity of particles
in the direction opposite to the mean temperature gradient.
A competition between two different phenomena, namely the
turbulent thermal diffusion and the turbulent diffusion
determines the conditions for the formation of large-scale particle
clusters in the vicinity of the mean temperature minimum.
The characteristic scale of the particle inhomogeneity formed due to
the turbulent thermal diffusion is much larger than the
integral scale of the turbulence. Furthermore, the characteristic
time scale of the formation of the particle inhomogeneity is much
longer than the characteristic turbulent time, i.e., this
is a mean-field effect.

The phenomenon of turbulent thermal diffusion has
been predicted theoretically in \cite{elperin_etal96}
and detected in different laboratory
experiments in stably and unstably temperature-stratified
turbulence \cite{EEKR04,BEE04,EEKR06a,eidelman_etal06,EKR10}.
This phenomenon is shown to be important for
atmospheric turbulence with temperature inversions
\cite{sofiev09} and for small-scale particle
clustering in temperature-stratified turbulence \cite{EKR10},
but it is also expected to be significant for
different kinds of heat exchangers, e.g., industrial boilers
where Reynolds numbers and temperature gradients are large.

In spite of the fact that turbulent thermal diffusion has
already been found in different types of laboratory
experiments and atmospheric flows, this effect has never
been observed in direct numerical simulations.
The main goal of this paper is
to find turbulent thermal diffusion
of non-inertial and inertial particles in direct
numerical simulations (DNS).

The paper is organized as follows.
In Sect.\ II we discuss the physics of the phenomenon
of turbulent thermal diffusion.
The numerical simulations for fluid, inertial
and non-inertial particles,
and the results of direct numerical simulations
are described in Sect.\ III.
Motions of inertial particles
are determined using a Lagrangian framework
(Sections III-B,C,D),
while non-inertial particles are described
using an Eulerian framework (Sect.\ III-E).
Conclusions are drawn in Sect.\ IV.

\section{Physics of turbulent thermal diffusion}

In this section we discuss the physics of
the phenomenon of turbulent thermal diffusion.
The number density $n_p(t,{\bm r})$ of particles advected
in a turbulent flow is determined by the following
equation \cite{CH43,AP81}
\begin{eqnarray}
{\partial n_p \over \partial t} + \bec{\nabla} {\bf \cdot}
\, (n_p \UUp - D \bec{\nabla} n_p) =0 ,
\label{B1}
\end{eqnarray}
where $D$ is the coefficient
of Brownian diffusion of
particles and $\UUp$ is a random velocity of
the particles which they acquire in a turbulent
velocity field $\UU$.
In order to study the formation of large-scale
inhomogeneous particle structures, Eq.~(\ref{B1})
for the particle number density is averaged over an
ensemble of turbulent velocity fields.
This yields an
equation for the mean number density, $N$, of particles
that has been derived using different approaches
in \cite{elperin_etal96,EKR00,PM02,RE05,sofiev09}:
\begin{eqnarray}
{\partial N \over \partial t} + \bec\nabla {\bf \cdot} \,
\big[N \, ({\bm W}_{\rm g} + {\bm V}^{\rm eff}) - (D+\DT) \,
\bec\nabla N \big] =0 \;,
\label{BB4}
\end{eqnarray}
where
\begin{eqnarray}
{\bm V}^{\rm eff} = - \tauf \, \overline{\UUp
\, (\bec\nabla \cdot \UUp)} =- \alpha \, \DT \,
\bec\nabla \ln\meanT
\label{BB5}
\end{eqnarray}
is the effective pumping velocity of particles
due to turbulent thermal diffusion,
$\meanT$ is the mean fluid temperature,
$\alpha$ is the turbulent thermal diffusion ratio (see below),
$\DT$ is the turbulent diffusion
coefficient, $\tauf$ is the fluid turbulent integral timescale,
overbars denote ensemble averaging,
${\bm W}_{\rm g} = \taup \, {\bm g}$  is the
terminal fall velocity
of particles,
${\bm g}$ is the gravitational acceleration,
$\taup$ is the Stokes time
that describes particle-fluid interactions.
Equation~(\ref{BB4}) is written for a zero mean
fluid velocity.
For large P\'eclet numbers, the coefficient
$\alpha$ is unity for non-inertial particles,
while for inertial particles $\alpha$ depends
on $\taup$, the Mach number and the
fluid Reynolds number \cite{elperin_etal96,sofiev09}.

Already simple
arguments allow us to estimate the effective particle
velocity ${\bm V}^{\rm eff}$.
Let us estimate the turbulent flux of particles
$\overline{n \, \UUp}$.
To this end we average Eq.~(\ref{B1})
over an ensemble of turbulent velocity fields, and subtract
the obtained averaged equation from Eq.~(\ref{B1}).
This yields an equation for the fluctuations of
the particle number density, $n=n_p - N$:
\begin{eqnarray}
{\partial n \over \partial t} + \bec\nabla {\bf \cdot}
\, \left[n \, \UUp - \overline{n \, \UUp} -
D \bec{\nabla} n\right] = - \bec{\nabla} {\bf \cdot}
\, (N \, \UUp) ,
\label{B2}
\end{eqnarray}
where for simplicity we consider the case of zero mean
particle velocity, $\overline{\UUp}=0$.
The left hand side of Eq.~(\ref{B2})
has dimension $n/\tau$, where the time $\tau$ can be
identified with the fluid turbulent integral timescale
$\tauf$ for large fluid Reynolds numbers.
The fluctuations of particle number density
can then be estimated as
\begin{eqnarray}
n \sim - \tauf  \, (\bec\nabla {\bf \cdot} \, \UUp) \, N
- \tauf \, (\UUp {\bf \cdot} \bec{\nabla}) N .
\label{B3}
\end{eqnarray}
Multiply Eq.~(\ref{B3}) by $\UUp$ and average this equation
over an ensemble of turbulent velocity fields, we arrive at a
formula for the turbulent flux of particles:
\begin{eqnarray}
\overline{n \, \UUp} = N \, {\bm V}^{\rm eff} - {\bm D_T} \,
\bec\nabla N ,
\label{B4}
\end{eqnarray}
where the first term on the right hand side of Eq.~(\ref{B4})
describes the turbulent flux of particles due to turbulent
thermal diffusion: $N \, {\bm V}^{\rm eff} = - N \, \tauf
\, \overline{\UUp \, (\bec\nabla \cdot \UUp)}$, and
the second term on the right hand side of Eq.~(\ref{B4})
determines the flux of particles caused by turbulent
diffusion (see, e.g., \cite{sofiev09}):
\begin{eqnarray}
{\bm D_T} \, \bec\nabla N = \tauf \,
\overline{(\UUp)_i \, (\UUp)_j} \, \nabla_j N \approx
\tauf \, \overline{U_i \, U_j} \, \nabla_j N ,
\label{TN3}
\end{eqnarray}
and ${\bm D_T} = \tauf \, \overline{U_i \, U_j}$ is the
turbulent diffusion tensor. For isotropic turbulence
${\bm D_T}=D_T \delta_{ij}$, where $D_T= \tauf \,
\overline{{\bm U}^2}/3$.

\subsection{Inertial particles}

The non-diffusive mean flux of particles,
$N \, {\bm V}^{\rm eff}$, toward the
mean temperature minimum is the main reason for
the formation of large-scale inhomogeneous distributions of
inertial particles in temperature-stratified turbulence.
Indeed, the steady-state solution
of Eqs.~(\ref{BB4}) and~(\ref{BB5})  for the
mean number density of inertial particles is given by
\EQ
\label{eq7}
\tilde N(z) = [\tilde T(z)]^{-\alpha}
\, \exp \Big[-\int_{z_0}^z \, [W_{\rm g} / \DT]
\, \dd z' \Big],
\EN
where $W_{\rm g}=|{\bm W}_{\rm g}|$ is the modulus of the terminal
fall velocity, $\tilde{T}=\meanT/\meanT_0$ is the non-dimensional
mean fluid temperature, $T_0$ is the mean temperature
far from the cooling zone, and $\tilde{N}=N/N_0$ is
the non-dimensional mean number density of particles.
Here the subscripts $0$ represent the values at the boundary,
and we neglect the small molecular flux of particles in
Eq.~(\ref{eq7}) which describes the Brownian diffusion
of particles.
Equation~(\ref{eq7}) implies that small particles are
accumulated in the vicinity of the mean temperature minimum.

In the following the mechanism of turbulent thermal
diffusion for inertial particles with material density,
$\rhop$, much larger than the fluid density, $\rho$,
will be explained \cite{elperin_etal96}.
The inertia causes particles inside the turbulent eddies
to drift out to the boundary regions between eddies.
This can be seen by considering the
equation of motion for particles, $\dd \UUp
/ \dd t =- (\UUp-\UU) / \taup$, where, for simplicity,
the gravity force is neglected.
The solution of the equation of motion for small
Stokes time reads \cite{M87}:
\begin{eqnarray}
\UUp = \UU - \taup \, {\dd \UU \over \dd t} + {\rm O}(\taup^2).
\label{TN1}
\end{eqnarray}
For large Reynolds numbers, this yields:
\begin{eqnarray}
\bec\nabla {\bf \cdot} \, \UUp = \bec\nabla {\bf \cdot} \, \UU
+ {\taup \over \rho}  \,\bec\nabla^2 p  + {\rm O}(\taup^2),
\label{TN2}
\end{eqnarray}
where $p$ is the fluid pressure.
For large P\'eclet numbers, when molecular diffusion of
particles in Eq.~(\ref{B1}) can be neglected,
it follows that $\bec\nabla {\bf \cdot} \, \UUp \propto
- \dd \ln n / \dd t$.
Therefore, in regions with maximum fluid pressure
(i.e., where $\bec\nabla^2 p < 0)$, there is accumulation
of inertial particles [i.e.,  $\dd n / \dd t \propto
- N \, (\taup /\meanrho) \,\bec\nabla^2 p > 0]$.
These regions have low vorticity,
high strain rate, and maximum fluid pressure.
Similarly, there is an outflow of inertial particles from
regions with minimum fluid pressure.

In homogeneous and isotropic turbulence without mean gradients
of temperature, a drift from regions with increased
concentration of particles by a turbulent flow is equiprobable
in all directions, and the pressure (temperature)
of the surrounding fluid is not correlated with the
turbulent velocity field.
There is only non-zero correlation $\overline{(\UU \cdot
\bec\nabla)  p}$ which contributes to the energy flux,
while $\overline{\UU p}=0$.

In temperature-stratified turbulence,
temperature and velocity fluctuations are correlated
due to a non-zero turbulent heat flux,
$\overline{\UU \, \theta}\not=\bm{0}$, where $\theta$
is the fluid temperature fluctuation.
Fluctuations of temperature cause pressure fluctuations,
which result in fluctuations of the number density
of particles.

\begin{table}[b]
\caption{Mechanism of particle transport $\overline{n \, \UU}$
for inertial particles along the direction of turbulent heat flux.}
\vspace{12pt}
\centerline {\begin{tabular}{c|c}
\multicolumn{2}{c}{$\xrightarrow{\hspace*{1.4cm}}\overline{\UU\theta}$} \\
\multicolumn{2}{c}{} \\
      {\sf{a}}        &          {\sf{b}}         \\
\hline
      $U_{{\rm p}x} > 0$        &     $U_{{\rm p}x} < 0$    \\
      $\theta > 0$        &     $\theta < 0$    \\
      $p > 0$        &     $p < 0$    \\
 $\bec\nabla{\bf \cdot} \, \UUp < 0\;$
&     $\;\bec\nabla{\bf \cdot} \, \UUp > 0$    \\
      $n > 0$        &     $n < 0$    \\
      $n \, U_{{\rm p}x}  > 0\;$        &     $\;n \, U_{{\rm p}x} > 0$\\
\multicolumn{2}{c}{} \\
\multicolumn{2}{c}{$\xrightarrow{\hspace*{1.4cm}}\overline{n \, \UUp}$}
\label{table_inert-particles}
\end{tabular}}
\end{table}

Increase of the pressure of the surrounding fluid is
accompanied by an accumulation of particles, and the
direction of the mean flux of particles coincides with that
of the turbulent heat flux.
The mean flux of particles is directed toward the
minimum of the mean temperature, and the particles
tend to be accumulated
in this region \cite{elperin_etal96,sofiev09,EKR00}.
To demonstrate that the directions of the mean flux of
particles and the turbulent heat flux coincide, we
assume that the mean temperature $T_2$ at point $2$
is larger than the mean
temperature $T_1$ at point $1$ (see, e.g., \cite{EEKR06a}).
We consider two small control volumes {``\sf{a}''} and {``\sf{b}''}
located between these two points;
see Table~\ref{table_inert-particles}.
Let the direction of the local turbulent
velocity in control volume {``\sf{a}''} at some instant
be the same as the direction of the turbulent heat flux
$\overline{\UU \, \theta}$
(i.e., along the $x$ axis toward point $1$)
and let the local turbulent velocity in control
volume {``\sf{b}''}, at the same instant, be directed opposite
to the turbulent heat flux (i.e., toward point $2$).

In temperature stratified turbulence, fluctuations of
pressure $p$ and velocity $\UU$ are correlated,
and regions with a higher level of pressure fluctuations
have higher temperature and velocity fluctuations.
The fluctuations of temperature $\theta$ and
pressure $p$ in control volume {``\sf{a}''} are positive
because $U_{{\rm p}x} \, \theta > 0$,
and negative in control volume {``\sf{b}''};
see Table~\ref{table_inert-particles}.
The fluctuations of particle number density $n$
are positive in control volume {``\sf{a}''} [because
particles are locally accumulated in the vicinity
of the maximum of pressure fluctuations, $\dd n / \dd t \propto
- N \, (\taup /\meanrho) \,\bec\nabla^2 p > 0$], and they are
negative in control volume {``\sf{b}''} (because there
is an outflow of particles from
regions with low pressure).
The flux of particles $n \, U_x$
is positive in control volume {``\sf{a}''} (i.e., it is
directed toward point $1$),
and it is also positive in control volume
{``\sf{b}''} (because both fluctuations of velocity and
number density of particles are negative
in control volume {``\sf{b}''}).
Therefore the mean flux of particles $\overline{n \, \UU}$
is directed, as is the turbulent heat flux
$\overline{\UU \, \theta}$, toward point~1.
This causes the formation of large-scale inhomogeneous
structures in the spatial distribution of inertial particles
in the vicinity of the mean temperature minimum.

\subsection{Non-inertial particles}

Let us now consider non-inertial particles advected by
the fluid flow such that the particle velocity coincides
with the fluid velocity, $\UUp=\UU$.
For low Mach numbers, using the anelastic approximation,
$\bec\nabla\cdot\UU \approx - \UU \cdot \bec\nabla
\ln \rho$, the effective velocity
of non-inertial particles is given by
\begin{eqnarray}
{V}_i^{\rm eff} = - \tauf \, \overline{U_i
\, (\bec\nabla \cdot \UU)} = \tauf \, \overline{U_i \, U_j} \,
\nabla_j \, \ln \meanrho ,
\label{B14a}
\end{eqnarray}
where $\meanrho$ is the mean fluid density.
When $\bec\nabla \, \overline{p} = \bm{0}$
and anisotropy of turbulence is weak,
the effective velocity of non-inertial particles is:
\begin{eqnarray}
{\bm V}^{\rm eff}= \DT \, \bec\nabla \, \ln \meanrho
= - \DT \,\bec\nabla \ln \meanT,
\label{B14}
\end{eqnarray}
where $\overline{p}$ is the mean fluid pressure.
The steady-state solution of
Eqs.~(\ref{BB4}) and~(\ref{B14}) for the
mean number density of non-inertial particles is given by
\begin{eqnarray}
\tilde N(z) = [\tilde \rho(z)]^{\DT /(D+\DT)} ,
\label{B15}
\end{eqnarray}
where $\tilde \rho(z)=\meanrho/\meanrho_0$,
where $\rho_0$ is the fluid density at the boundary.

It follows from Eq.~(\ref{B14}) that the non-diffusive
flux of particles is directed toward the maximum
of the fluid density.
The physics of the accumulation of the non-inertial
particles in the vicinity of the maximum of the mean
fluid density (or the minimum of the mean fluid
temperature) is quite different from that of inertial
particles and can be explained as follows
(see Table~\ref{table_non-particles}).
Let us assume that the mean fluid density $\meanrho_2$
at point $2$ is larger than the mean
fluid density $\meanrho_1$ at point $1$.
Consider two small control volumes {``\sf{a}''} and {``\sf{b}''}
located between these two points, and let the direction
of the local turbulent velocity in control volume
{``\sf{a}''} at some instant be the same
as the direction of the mean fluid density gradient
$\bec\nabla \, \meanrho$
(i.e., along the $x$ axis toward point $2$).
Let the local turbulent velocity in control volume
{``\sf{b}''} be directed at this instant opposite to the mean
fluid density gradient (i.e., toward point $1$).

In a fluid flow with an imposed mean temperature gradient
(i.e., an imposed mean fluid density gradient),
one of the sources of particle number density fluctuations,
$n \propto - \tauf \, N \, (\bec\nabla {\bf \cdot} \, \UU)$,
is caused by the non-zero $\bec\nabla\cdot\UU \approx
- \UU \cdot \bec\nabla \ln \rho \not=0$
[see the first term on the right
hand side of Eq.~(\ref{B3})].
Since the fluctuations of the fluid velocity $\UU$
are positive in control volume {``\sf{a}''} and negative in
control volume {``\sf{b}''}, we have $\bec\nabla
{\bf \cdot} \, \UU < 0$ in control volume {``\sf{a}''},
and $\bec\nabla {\bf \cdot}
\, \UU > 0$ in control volume {``\sf{b}''}.
Therefore, the fluctuations of the particle number
density $n \propto - \tauf \, N \, (\bec\nabla
{\bf \cdot} \, \UU)$
are positive in control volume {``\sf{a}''} and negative
in control volume {``\sf{b}''}.
However, the flux of particles $n\, U_x$ is positive
in control volume {``\sf{a}''} (i.e., it is directed to
point $2$), and it is also positive in control
volume {``\sf{b}''} (because both fluctuations of fluid
velocity and number density of particles are negative
in control volume {``\sf{b}''}).
Therefore, the mean flux of particles $\overline{n\, \UU}$
is directed, as is the mean fluid density gradient
$\bec\nabla \, \meanrho$, toward point~2.
This forms large-scale heterogeneous
structures of non-inertial particles
in regions with a mean fluid density maximum.

\begin{table}
\caption{Mechanism of particle transport $\overline{n \, \UU}$
for non-inertial particles along the direction
of the mean fluid density gradient.}
\vspace{12pt}
\centerline {\begin{tabular}{c|c}
\multicolumn{2}{c}{$\xrightarrow{\hspace*{1.4cm}}\nab\meanrho$} \\
\multicolumn{2}{c}{} \\
      a        &          b         \\
\hline
      $U_x > 0$        &     $U_x < 0$    \\
 $\bec\nabla{\bf \cdot} \, \UU < 0\;$
&     $\;\bec\nabla{\bf \cdot} \, \UU > 0$    \\
      $n > 0$        &     $n < 0$    \\
      $n \, U_x  > 0\;$        &     $\; n \, U_x > 0$ \\
\multicolumn{2}{c}{} \\
\multicolumn{2}{c}{$\xrightarrow{\hspace*{1.4cm}}\overline{n \, \UU}$}
\label{table_non-particles}
\end{tabular}}
\end{table}

\section{Direct numerical simulations}

\subsection{DNS model for the fluid}

In this study,
equations for the fluid are solved by employing DNS in
an Eulerian framework.
All numerical simulations have been performed using the
{\sc Pencil Code} \cite{PC} (for details of the code,
see Refs.~\cite{Johansen07,Haugen_kragset10}).
The set of compressible hydrodynamic equations is solved for
the fluid density $\rho$, the fluid velocity $\bm{U}$,
and the specific entropy $s$:
\begin{eqnarray}
&& \frac{\DD \ln \, \rho}{\DD t} = - \bec{\nabla} {\bf \cdot} \, {\bm U},
\label{density}\\
&&\frac{\DD \bm{U}}{\DD t} = - \frac{1}{\rho}\left[\bec{\nabla} p
- \bec{\nabla} {\bf \cdot}(2 \rho \nu \bm{\mathsf{S}})\right] + {\bm f},
\label{velocity}\\
&& T\frac{\DD s}{\DD t} = \frac{1}{\rho} \bm{\nabla} \cdot K \bm{\nabla}T
+ 2 \nu \bm{\mathsf{S}}^2 - c_{\rm P} (T - T_{\rm ref}),
\label{entropy}
\end{eqnarray}
where $T=T_0\exp(s/c_{\rm V})(\rho/\rho_0)^{\gamma-1}$ is the fluid temperature,
$\gamma=c_{\rm P}/c_{\rm V}$ is the ratio of specific heats
at constant pressure and constant volume, respectively,
$\DD/\DD t=\partial/\partial t +\UU\cdot\nab$ is the advective derivative,
${\bm f}$ is the external forcing function,
$p=(c_{\rm P}-c_{\rm V})\rho T$ is the fluid pressure,
$\nu$ is the kinematic viscosity,
and $K$ is the thermal conductivity.
The traceless rate of strain tensor is given by
\begin{eqnarray*}
\SSS_{ij}=\frac{1}{2}(U_{i,j} + U_{j,i})-\frac{1}{3}\delta_{ij}\nab\cdot\UU,
\end{eqnarray*}
where $U_{i,j}=\nabla_j U_i$.
The last term in the entropy equation~(\ref{entropy})
determines the cooling,
which causes a temperature minimum at $z=0$, where
\begin{eqnarray*}
T_{\rm ref} = T_0 + \delta T \, \exp (- z^2 / 2 \sigma^2),
\end{eqnarray*}
$z$ is the vertical coordinate.

The size of the simulation domain is $L$ in all three directions.
The smallest wavenumber in the domain, $k_1=2\pi/L$.
Wavenumbers are measured in units of $k_1$,
and lengths in units of $k_1^{-1}$.
The width of the cooling function is $\sigma=0.5/k_1$.
Furthermore, the
strength of the cooling function, $\delta T$,
is such that the relative
contrast between temperature maximum and temperature minimum is $\sim1.6$.
In all simulations the Prandtl number ${\rm Pr}=\nu / K=1$.

Turbulence in the simulation box is
produced by the forcing function ${\bm f}$,
which is solenoidal and non-helical, i.e.,
$\bec{\nabla}\cdot{\bm f}={\bm f}\cdot\bec{\nabla}\times{\bm f}=0$,
and injects energy and momentum
perpendicular to a random wavevector
whose direction changes every timestep, but
its length is approximately $\kf$ (see \cite{HB06}).
The strength of the forcing is set such
that the maximum Mach number is below 0.5.
It has been checked that decreasing the Mach number
by a factor of 5 does not have a significant
effect on the results.

\subsection{Lagrangian model for inertial particles}

The equation of motion of the inertial particles
is solved numerically in a Lagrangian framework.
Particles are treated as point particles and we
consider the one-way coupling approximation, i.e., there is an
effect of the fluid on the particles only, while the
particles do not influence the fluid motions.
This is a good approximation
when the spatial density of particles is much smaller than
the fluid density.
The particle equation of motion is
\begin{equation}
{\dd \UUp \over \dd t} = {\bm g} - {\UUp-\UU \over \taup},
\label{particles}
\end{equation}
where $\dd\bm{X}/{\dd t}=\UUp$,
\begin{eqnarray*}
\taup={m_p \over 3 \pi \, \mu \,d \, (1-f_{\rm c})}
\end{eqnarray*}
is the Stokes time,
$m_p = \rhop \, (\pi d^3 / 6)$ is the particle mass
of the spherical form,
$d$ is the particle diameter, $\rhop$
is the particle material density, $\mu=\rho \, \nu$
is the dynamic viscosity,
$f_{\rm c}=0.15\Rey_{\rm p}^{0.687}$~\cite{CST11},
and $\Rey_{\rm p}=|\UUp-\UU|d/\nu$ is
the particle Reynolds number.
The key parameter of the problem is the Stokes number
$\St=\taup/\tauk$ that is based on the particle Stokes
time and the Kolmogorov timescale $\tauk=\tauf/\sqrt{\Rey}$,
where $\tauf=1/u_{\rm rms}\kf$ is the turbulent integral
timescale, $\Rey=u_{\rm rms}/\nu\kf$ is the fluid Reynolds number
and $u_{\rm rms}$ is the root mean square fluid velocity.
The parameter $f_{\rm c}$ is used
for inertial particles with $\Rey_{\rm p} \geq 1$.
Even though $f_{\rm c}$ is indeed used for calculating the particle
forces in the simulations it is set to zero when calculating
the Stokes number of a given particle.
This is justified by the fact that for all particles
considered here $f_{\rm c}$ is always very small and, furthermore, the Stokes number is only used for illustrative purposes.
The ratio between the particle material density and the fluid
mass density is $S=\rhop/\rho_0$, and for all simulations
$S=1000$ outside the cooled zone.

\begin{figure}
\centering
\includegraphics[width=0.5\textwidth]{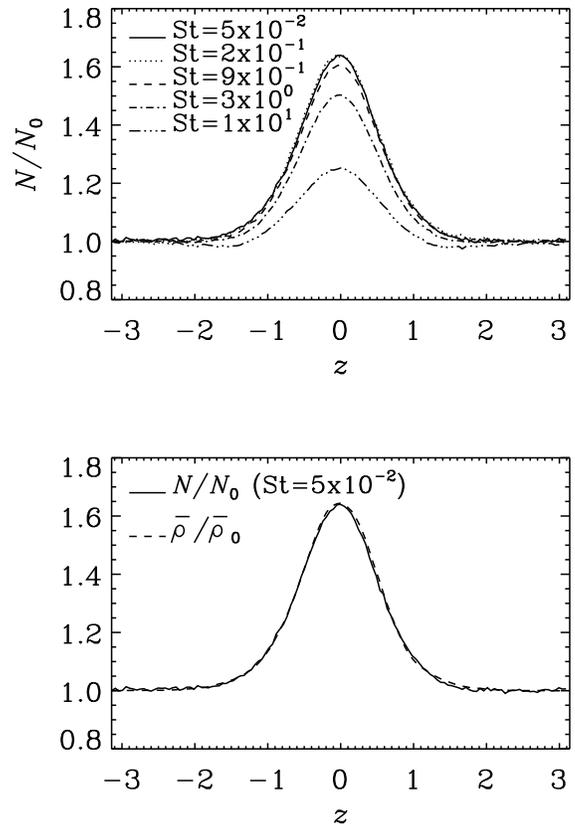}
\caption{Upper plot: vertical profile of the mean
number density of
inertial particles for different Stokes numbers
for simulations with
$\kf=5k_1$, $\Rey=240$ in the case where the kinematic
viscosity is kept constant and no gravity.
Here the Stokes number $\St$ is based on the fluid density and temperature at the boundary.
Lower plot: same as the upper
plot, but showing the particle number density only
for the smallest Stokes
numbers together with the fluid mass density.}
\label{number_density_profiles}
\end{figure}

\begin{figure}
\centering
\includegraphics[width=0.5\textwidth]{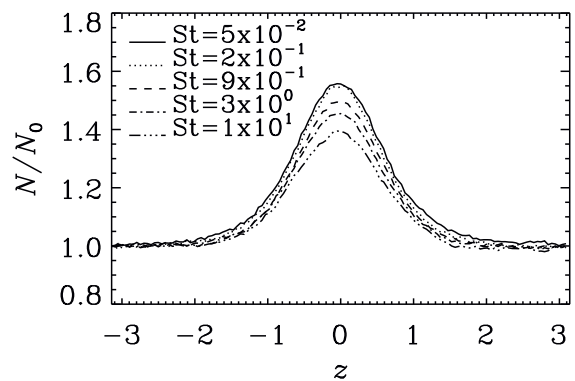}
\caption{The same as the upper plot of \Fig{number_density_profiles}
but for the case where the
dynamic viscosity is $\mu= \mu_* \sqrt{T(z)/T_0}$,
$\mu_*$ is kept constant,
and the Stokes number $\St$ is based on the fluid density and temperature at the boundary.
\label{number_density_profiles_temp}}
\end{figure}

To demonstrate robustness of the effect of particle
accumulation in the vicinity of mean temperature minimum we use three
different models for the dynamic viscosity $\mu$:
(i) $\mu= \mu_\ast \sqrt{T(z)/T_0}$, where $\mu_\ast=$const;
(ii) $\mu$=const; (iii) $\mu= \rho(z) \, \nu$,
where $\nu=$const.

For most of the simulations the number of
Eulerian grid points are $128^3$, while the number of Lagrangian
particles is $5\times 10^5$.
It has been verified that
increasing the number of grid points
up to $512^3$ or the number of particles
up to $10^7$ has no effect on the results.

In simulations without gravity, periodic boundary
conditions are used in all
directions both for the fluid and for the particles.
When gravity is taken into account in the simulations,
particles are made elastically
reflecting from the vertical boundaries.
It has been checked that adding small Brownian diffusion
of particles does not affect the results.

In all our simulations,
gravity is ignored in \Eq{velocity} for
the fluid, but for the simulations presented
in \Sec{secC} it is included
in \Eq{particles} for particle motions.
This situation
has direct applications to atmospheric turbulence with
lower-troposphere temperature inversions, where the density
scale height due to gravity is about 8~km,
while the characteristic temperature inhomogeneity scale
inside the temperature inversions is about 500--800~m,
and the integral scale of turbulence is about 50--100~m
(see, e.g., \cite{BLA97}).

\subsection{Inertial particles in the absence of gravity}
\label{secC}

Let us first discuss the results of the numerical
simulations concerning the formation of the
inertial particle inhomogeneities
in the absence of gravity.
In Figs.~\ref{number_density_profiles}
and~\ref{number_density_profiles_temp} we plot the vertical profile
of the mean number density of particles
for simulations with $\kf=5k_1$, $\Rey=240$,
different Stokes numbers, and different formulations for the dynamic viscosity.
Inspection of these figures
shows that, for a wide range of Stokes numbers, the maximum
in the particle number density is located in the vicinity
of the mean temperature minimum.

\begin{figure}
\centering
\includegraphics[width=0.5\textwidth]{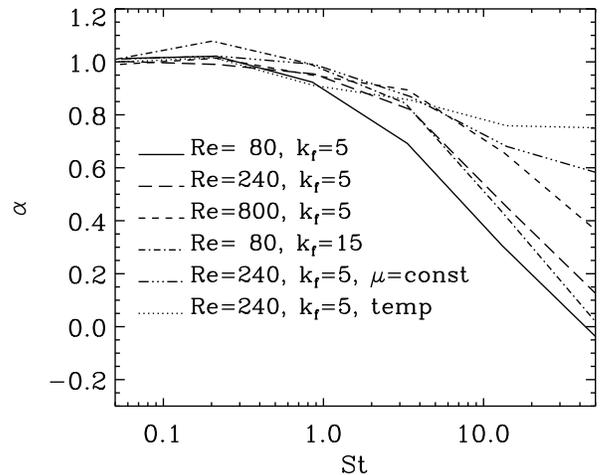}
\caption{ The parameter $\alpha$ versus the Stokes number $\St$ for
  different Reynolds numbers and forcing wavenumbers in the absence of
  gravity.  For the dashed-triple-dotted line the dynamic viscosity,
  $\mu$, is kept constant while for the dotted line named
  ``temp'' the dynamic viscosity $\mu= \mu_* \sqrt{T(z)/T_0}$,
 where $\mu_*$ is kept constant.
 For all other lines the dynamic viscosity,
 $\mu=\rho(z) \nu$, where the kinematic viscosity $\nu$ is
 kept constant.
Here the Stokes number $\St$ is based on the values
of fluid density and temperature at the boundary.
\label{alpha}}
\end{figure}

In the absence of gravity ($W_{\rm g}=0$),
the parameter $\alpha$ is given by
$\alpha= -\ln\tilde{N} / \ln\tilde{T}$; see Eq.~(\ref{eq7}).
In \Fig{alpha} the Stokes number dependence of the parameter
$\alpha$ is shown for different Reynolds numbers and different
forcing scales. It can be seen that the parameter $\alpha$
reaches its maximum value
for $\St_*\approx 0.2$, while for Stokes numbers
$\St>\St_*$, the value of $\alpha$ decreases monotonically.
A critical Stokes number, $\St_{\rm c}$, separating the large and
small Stokes number regimes might be defined such that
$\alpha \geq 1$ for $\St < \St_{\rm c}$,
while for $\St > \St_{\rm c}$ the parameter
$\alpha$ is decreasing steeply with
increasing Stokes number.

Since in the simulations the rms Mach number
is not very small (around 0.2),
the maximum value of the parameter $\alpha$
is only slightly larger than 1.
Indeed, the contribution of particle inertia to
the coefficient $\alpha$ for small Stokes numbers is
determined by $\bec\nabla {\bf \cdot} \, \UUp
\propto (\taup / \overline{\rho})  \,\bec\nabla^2 p$.
On the other hand,
$\bec\nabla^2 p / \overline{\rho} \propto c_{\rm s}^2 \, \bec\nabla^2
\theta / \overline{T} \propto M^{-2} \, u_{\rm rms}^2 \,
\bec\nabla^2 \theta / \overline{T}$,
where $M=u_{\rm rms} /c_{\rm s}$ is the Mach number,
and $c_{\rm s}$ is the speed of sound.
The deviation of the parameter $\alpha$
from 1 is of the order of \cite{EKR10}
\begin{equation}
\alpha - 1 \propto {\taup \over \tauf} \, M^{-2}.
\label{alp1}
\end{equation}
In this estimate we took into account that
$|\overline{\UU \, \bec\nabla^2 \theta}| \sim \ell^{-2}
|\overline{\UU \, \theta}|$ and the turbulent heat flux is
$\overline{\UU \, \theta} =
- \kappa_{_{T}} \bec\nabla \overline{T}$, where
$\kappa_{_{T}} \sim \tauf \, u_{\rm rms}^2$
and $\ell = \tauf \, u_{\rm rms}$.

Therefore, decreasing the Mach number causes $\alpha$
to increase due to particle inertia.
For instance, in atmospheric turbulence the Mach number
is of the order of $(0.1 - 3) \times 10^{-3}$ and $\alpha$
is about 10 \cite{sofiev09}.
On the other hand, the observed effect of accumulation
of inertial particles occurs
only for turbulent flows with large fluid Reynolds number.
Decreasing the Mach number in the {\sc Pencil Code}
will either result in a shorter time step or a lower
Reynolds number.
This implies that we cannot easily reach at the same
time small Mach numbers and large fluid Reynolds number.
Therefore, the observations of the accumulation of inertial
particles in the parameter range $\alpha\gg1$
is not currently possible with the {\sc Pencil Code}.

The parameter $\alpha$ for the heavy particles
decreases strongly with $\taup\propto \St$ (see \Fig{alpha}
in the large Stokes number regime).
Furthermore, it is expected that,
in the large Stokes number regime, $\alpha$ is
independent of $\Rey$ for a particle as long
as $u_{\rm rms}$ and $\kf$ are kept constant.
This is due to the fact that particles with
$\taup \sim \tauf$ are affected primarily by
the largest turbulent eddies in the flow.
This is because the smaller turbulent eddies
have turnover times much shorter than the particle's
Stokes time, and they cannot accelerate the particles.
Since the Stokes number is based on the Kolmogorov
time $\tauk$, this implies that
$\St_{\rm c} \propto \sqrt{\Rey}$,
which is indeed what is found in \Fig{alpha}.

When experiencing a high fluid density, a
particle in the large Stokes number regime will more easily be accelerated
by turbulence if the kinematic viscosity $\nu$ is constant
than if the dynamic viscosity $\mu=\rho\nu$ is constant.
This is due to the fact that for constant $\nu$ a high fluid density is
associated with a smaller Stokes number, while for constant $\mu$ the
Stokes number is independent of the fluid density.
Due to this it is expected that
particles with large Stokes numbers tend to be more depleted from the high
density regions when $\nu$ is constant than when $\mu$ is constant.
This explains why the simulation shown in \Fig{alpha} with constant
$\mu$ has a much shallower fall-off with increasing Stokes number
in the large Stokes number regime than
the simulations with constant $\nu$.

If the integral scale of turbulence is close to the scale
of the thermal cooling layer, a particle trapped inside a
turbulent eddy might travel
across the whole cooling zone during one eddy turnover time.
This will effectively smear out the peak of the particle
number density, and consequently also
decrease the parameter $\alpha$.
For simulations with $\kf=5k_1$, the integral scale
is comparable to the cooling scale, and the peak of $\alpha$ for $\kf=5k_1$
is lower than for simulations with $\kf=15k_1$ (see \Fig{alpha}).
This trend is expected to continue for yet larger $\kf$.

\begin{table}
\caption{The coefficient $\alpha$, the relative maximum of the
 particle number density and profile width for simulations with
 $\kf=5k_1$, $\Rey=240$ and for different
 formulations of the dynamic viscosity: (i) in the model $f_1$ the dynamic viscosity is $\mu= \mu_* \sqrt{T(z)/T_0}$,
 where $\mu_*$ is constant; (ii) in the model $f_2$ the dynamic viscosity is constant;  (iii) in the model  $f_3$
 the dynamic viscosity $\mu=\rho(z) \, \nu$, where the kinematic viscosity $\nu$ is constant.
 The profile width $L_N$ is given at the level
 when the number density is decreased in $e$ times.
 }
\vspace{12pt}
\centerline {\begin{tabular}{c|ccc|ccc|ccc}
      &\multicolumn{3}{c}{$\alpha$}&\multicolumn{3}{c}{$N_{\rm max}/N_0$}&\multicolumn{3}{c}{Profile width}\\
$\St$ &  $f_1$&  $f_2$&   $f_3$&  $f_1$&  $f_2$&   $f_3$ &   $f_1$ &   $f_2$ &  $f_3$ \\
\hline
0.1   & 1.00  & 1.01  & 1.00   &  1.55 &  1.66 &   1.63 &    0.88 &    0.80 &   0.79 \\
1     & 0.91  & 0.98  & 0.95   &  1.49 &  1.62 &   1.60 &    0.86 &    0.80 &   0.79 \\
10    & 0.80  & 0.75  & 0.59   &  1.42 &  1.44 &   1.35 &    0.90 &    0.82 &   0.81 \\
14    & 0.76  & 0.68  & 0.45   &  1.39 &  1.39 &   1.25 &    0.90 &    0.83 &   0.82 \\
\label{table_ntilde}
\end{tabular}}
\end{table}

In addition, \Fig{alpha} shows that the difference in the
parameter $\alpha$ determined for different formulations
of the dynamic viscosity $\mu$, is very small
for $\St \ll 1$, and it is small for $\St \sim 1$.
The real difference in the parameter $\alpha$ is only
observed for large
Stokes numbers, $\St > 10$ (see TABLE~\ref{table_ntilde}).
In this case, the largest value of $\alpha$ occurs
for the dynamic viscosity $\mu= \mu_\ast
\sqrt{T(z)/T_0}$, where $\mu_\ast=$const.
On the other hand, the difference in relative maximum of the
particle number density and the profile width for different formulations
of dynamic viscosity is very minor (see TABLE~\ref{table_ntilde}).
Also scale-separation increases effect of the particle
accumulation for $\St > 1$ [see \Fig{alpha}].
However, using different parameters
and different formulations of the dynamic viscosity,
we always observe the effect of particle
accumulation in the vicinity of the mean temperature
minimum. This is an indication of the robustness
of this phenomenon.

\subsection{Inertial particles with gravity}

\begin{figure}
\centering
\includegraphics[width=0.45\textwidth]{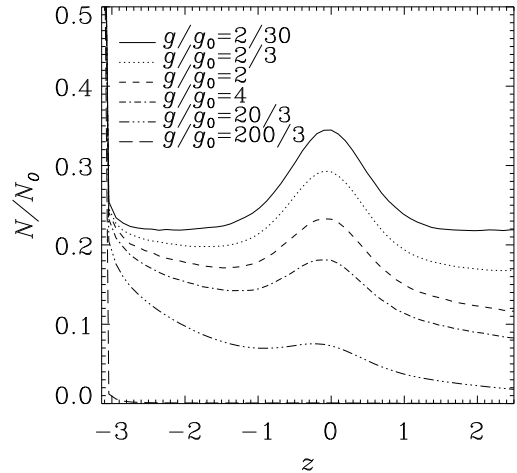}
\caption{Vertical profile of the mean number density for different
gravitational accelerations, $\St=1$ and $\kf=5k_1$,
where the kinematic viscosity $\nu$ is kept constant.}
\label{fig3}
\end{figure}

For heavy inertial particles with large Stokes numbers,
gravity plays a crucial role, and must be
included in the simulations.
In \Fig{fig3}, the mean particle number density is shown as a function
of vertical position $z$ for simulations with
$\St=1$, $\kf=5k_1$,
and different gravitational accelerations.

In the following, gravity is measured in the units of
$g_0=\DT/(\tauk L)$, so that $g/g_0 = L/\Lg$
for a particle of Stokes number unity.
Here, $\Lg=\DT/W_{\rm g}$ is the characteristic
scale of the mean particle number
density variations
due to the gravity for an isothermal
case, $\DT=u_{\rm rms}/3\kf$
is the turbulent diffusion
coefficient,  and $L$ is the height of the box.

As the particle sedimentation velocity is increased,
the particle number density profile is more
and more tilted, as expected.
For large particle sedimentation velocity $(g/g_0=200/3)$,
almost all particles have accumulated at the lower wall
of the box (see \Fig{fig3}).

\begin{figure}
\centering
\includegraphics[width=0.45\textwidth]{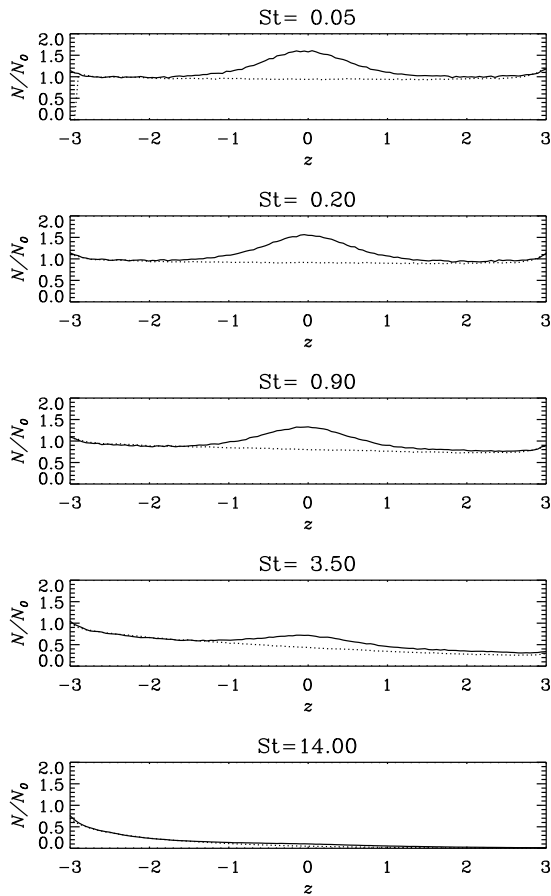}
\caption{
Vertical profile of the mean number density (solid line)
for $g/g_0=2/3$, $\kf=5k_1$ and different Stokes numbers
which are based on the fluid density and temperature
at the boundary.
Here the kinematic viscosity $\nu$ is kept constant.
The dotted line represent the isothermal reference case.
}\label{fig4}
\end{figure}

In \Fig{fig4}, the vertical profile of the particle number density
is shown for $g/g_0=2/3$ and different Stokes numbers.
By fitting these results with \Eq{eq7},
the parameter $\alpha$ is found to be about 1 for the three smallest
Stokes numbers in \Fig{fig4}, while for $\St=3.5$
and $\St=14$ a best fit of $\alpha$
is found around 0.8 and 0.5, respectively.
For large Stokes numbers the maximum particle
number density is found near the bottom wall
of the box due to the large sedimentation velocity.

\subsection{Non-inertial particles}

\begin{figure}
\centering
\includegraphics[width=0.5\textwidth]{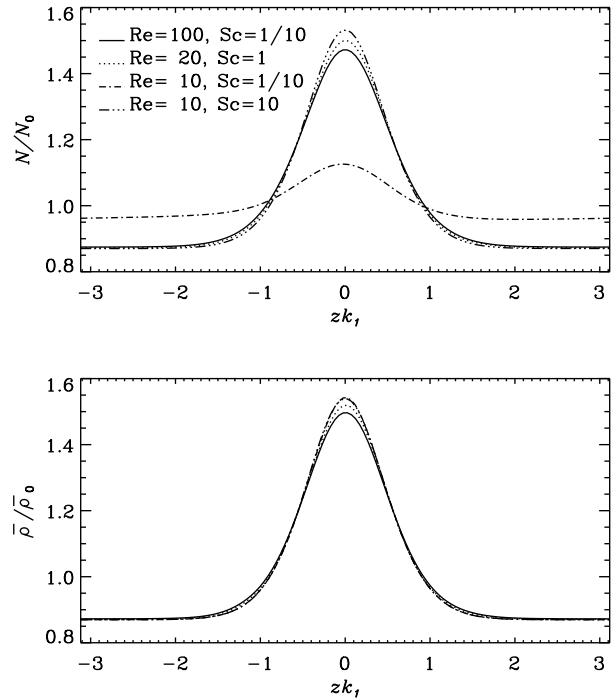}
\caption{Vertical profiles of the mean number density
  (upper panel) and the
  mean fluid density (lower panel) of
  non-inertial particles
  for simulations with $\kf=5k_1$ and different Reynolds
  and Schmidt numbers.}
\label{fig5}
\end{figure}

For comparison with the numerical simulations performed
for inertial particles in a Lagrangian framework,
we describe in this section numerical simulations
for non-inertial particles, where the equations for both,
fluid and particles are solved by employing DNS in
an Eulerian framework.
In particular, we now solve Eq.~(\ref{B1})
for the number density of
non-inertial particles $n_p$ with $\bm{U}_p=\bm{U}$, and
the fluid density $\rho$, the fluid velocity $\bm{U}$,
and the specific entropy $s$, using again the
{\sc Pencil Code} \cite{PC}.
For these simulations the resolution is $128^3$.
For the fluid we apply the same conditions as was described
in previous subsections.
In particular, the periodic
boundary conditions are used in these simulations in three
directions for Eqs.~(\ref{B1})
and (\ref{density})--(\ref{entropy}).

Non-inertial particles are characterized by the following
dimensionless parameters: ${\rm Pe}=u_{\rm rms}/D\kf$
is the P\'eclet number and ${\rm Sc}=\nu/D$
is the Schmidt number.
The results of the simulations, shown in \Fig{fig5},
demonstrate the accumulation of non-inertial
particles in the vicinity of the maximum of the mean
fluid density (or the minimum of the mean fluid temperature).
This is in agreement with the theoretical predictions
\cite{elperin_etal96}.

The results shown in \Fig{fig5} demonstrate that
the profiles of
the mean fluid density and the mean particle number density
are similar in numerical simulations with large
P\'eclet and Reynolds numbers.
However, when the P\'eclet number,
${\rm Pe}= {\rm Re} \, {\rm Sc}$
is not large, the molecular diffusion $D$
becomes important, and the profiles of the mean fluid
density and the mean particle number density are,
according to Eq.~(\ref{B15}), no longer similar.

\section{Conclusions}

This study is the first numerical demonstration of the
existence of the phenomenon of turbulent thermal diffusion
of inertial and non-inertial particles
in forced, temperature stratified turbulence.
The inertial particles are described using a Lagrangian
framework, while non-inertial particles and the fluid flow
are determined using an Eulerian framework.
The phenomenon of turbulent thermal diffusion has been
studied for different Stokes and fluid Reynolds numbers,
P\'eclet numbers as well as different forcing scales
of the turbulence.
Furthermore, the effect of gravity has been included in
simulations.

In all simulations, with different parameters
and different formulations of the dynamic viscosity,
we always observe the effect of particle
accumulation in the vicinity of the mean temperature
minimum due to turbulent
thermal diffusion for $\St<1$. This effect is robust and
the results of the numerical simulations are
in agreement with theoretical studies \cite{elperin_etal96,EKR00},
laboratory experiments \cite{BEE04,eidelman_etal06,EKR10}
and atmospheric observations \cite{sofiev09}.
When $\St>1$, this effect is decreasing with Stokes number.

\begin{acknowledgements}
This work was supported in part by the Norwegian Research Council
project PAFFrx (186933) (NELH), by COST Action MP0806 (IR),
and by the European Research Council under the
AstroDyn Research Project 227952 (AB).
The authors acknowledge the hospitality of NORDITA.

\end{acknowledgements}

\end{document}